\documentclass[]{sig-alternate}

\makeatletter
\def\@copyrightspace{\relax}
\makeatother

\usepackage{dirtree}
\usepackage{color}
\usepackage[hidelinks]{hyperref}
\usepackage{xspace}
\definecolor{light-gray}{gray}{0.9}
\usepackage{subcaption}
\usepackage[numbers]{natbib} 
\usepackage{qtree}
\usepackage{booktabs}

\usepackage[version=0.96]{pgf}
\usepackage{tikz}
\usetikzlibrary{arrows,shapes,snakes,automata,backgrounds,petri}
\usepackage{multirow}
\usepackage{multicol}
\usepackage{balance}

\usepackage{microtype}
\usepackage{paralist}

\newlength{\emstr}
\newcommand{\boldpara}[1]{%
  \smallskip%
  \par\noindent\textbf{\textit{#1}}\hspace{\emstr}
}%

\usepackage{listings}
\newcommand{\cut}[1]{}

\newcommand{\etal}{\hbox{\emph{et al.}}\xspace}
\newcommand{\eg}{\hbox{\emph{e.g.}}\xspace}
\newcommand{\ie}{\hbox{\emph{i.e.}}\xspace}

\newcommand{\Ngram}{$N$-gram\xspace}
\newcommand{\ngrams}{$n$-grams\xspace}
\newcommand{\haggis}{\textsc{Haggis}\xspace}
\newcommand{\deckard}{\textsc{Deckard}\xspace}
\newcommand{\libraryDset}{\textsc{Library}\xspace}
\newcommand{\projectsDset}{\textsc{Projects}\xspace}
\newcommand{\problem}{CFG extension problem\xspace}

\newcommand{\Data}{\mathcal{D}}
\newcommand{\calT}{\mathcal{T}}
\newcommand{\Id}{\mathcal{I}}
\newcommand{\setF}{\mathcal{F}}
\newcommand{\R}{\mathbb{R}}
\newcommand{\Prob}[1]{\Pr \lbrack #1 \rbrack}
\newcommand{\join}{\text{\tiny join}}
\newcommand{\post}{\text{\tiny post}}

\newcommand{\ML}{\text{\tiny ML}}
\newcommand{\geom}{\text{\tiny geom}}

\begin{document}
%
\conferenceinfo{FSE}{'14 Hong Kong, China}

\title{Mining Idioms from Source Code}

\numberofauthors{1} 
%
\author{
%
%
\alignauthor
Miltiadis Allamanis,  Charles Sutton\\
       \affaddr{School of Informatics, University of Edinburgh}\\
       \affaddr{Edinburgh EH8 9AB, UK}\\
       \email{\{m.allamanis,csutton\}@ed.ac.uk}}
\date{30 July 1999}

\maketitle\vspace*{-2.5em}
\begin{abstract}
We present the first method for automatically mining code idioms
from a corpus of previously written, idiomatic software projects.
We take the view that a \emph{code idiom} is a syntactic fragment
that recurs across projects and has a single semantic role.
Idioms may have metavariables, such as the body of a \lstinline|for| loop.
Modern IDEs commonly provide facilities for manually defining idioms 
and inserting them on demand, but this does not help programmers to
write idiomatic code in languages or using libraries with which they are unfamiliar.
We present \haggis, a system for mining code idioms that
builds on recent advanced techniques from statistical
natural language processing, namely, nonparametric Bayesian probabilistic
tree substitution grammars.  We apply \haggis to several of the most popular
open source projects from GitHub. We present a wide range of evidence
that the resulting idioms are semantically meaningful, 
demonstrating that they do indeed recur across software projects
and that they occur more frequently in illustrative code examples
collected from a Q\&A site. 
Manual examination of the most common idioms indicate
that they describe important program concepts, including object creation,
exception handling, and resource management.
  \end{abstract}




\section{Introduction}
Programming language text is a means of human communication.
Programmers write code not simply to be executed by a computer,
but also to communicate the precise details of the code's operation
to later developers who will adapt, update, test and maintain the code. 
It is perhaps for this reason that source code is
\emph{natural} in the sense described by \citet{hindle2012naturalness}.
Programmers themselves use the term \emph{idiomatic} to refer to code
that is written in a manner that other experienced developers
find natural.  Programmers believe that it is important to write idiomatic code.
This is evidenced simply by the amount of time that programmers spend
telling other programmers how to do this.
For example, Wikibooks has a book devoted to C++ idioms \cite{cppidioms},
and similar guides are available for Java \cite{javaidioms} and JavaScript \cite{jspatterns,idiomaticjs}.  A guide on GitHub
for writing idiomatic JavaScript \cite{idiomaticjs} has more
6,644 stars and 877 forks.  
A search for the keyword ``idiomatic'' on StackOverflow yields
over 49,000 hits; all but one of the first 100 hits are
 questions about what the {idiomatic} method is for performing a given task.

The notion of \emph{code idiom} is one that is commonly used but seldom
defined.  We take the view that {an idiom is a syntactic fragment
that recurs frequently across software projects and has a single semantic role.}
Idioms may have metavariables that abstract over identifier names
and code blocks.
For example, in Java the loop
\lstinline|for(int i=0;i<n;i++) { ... }|
is a common idiom for iterating over an array.
It is possible to express this operation in many other ways, such
as a \lstinline{do-while} loop or using recursion, but 
as experienced Java programmers ourselves, we would find this alien and more
difficult to understand.
Idioms differ significantly from previous notions of textual patterns
in software, such as code clones \cite{roy2007survey} and API patterns \cite{zhong2009mapo}.  Unlike clones, idioms commonly recur across projects,
even ones from different domains, and unlike API patterns, idioms 
commonly involve syntactic constructs, such as iteration and exception handling.
A large number of example idioms, all of which are automatically identified by our system,
are shown in Figures~\ref{fig:sampleIdioms} and~\ref{fig:snippets}.

Major IDEs currently support idioms by including features that allow programmers to define
idioms and easily reuse them. Eclipse's 
SnipMatch \cite{eclipseSnipMatch}
and IntelliJ IDEA live templates \cite{jetbrainsLiveTemplates}
allow the user to define custom snippets of code that can be inserted on demand.
NetBeans includes a similar ``Code Templates''
feature in its editor. Recently, Microsoft created
Bing Code Search \cite{bingCodeSearch}
that allows users to search and add snippets to their code, by retrieving
code from popular coding websites, such as StackOverflow. The fact that
all major IDEs include features that allow programmers to manually define and use
idioms attests to their importance.

We are unaware, however, of methods for \emph{automatically} identifying 
code idioms. This is a major gap in current tooling for software development,
which causes significant problems.
First, software developers cannot use manual IDE tools for idioms
without significant effort to organize the idioms of interest and 
then to manually add them to the tool.  This is especially an obstacle
for less experienced programmers that do not know which idioms they should
be using. Indeed, as we demonstrate later, many idioms are library-specific,
so even an experienced programmer will not be familiar with the code idioms
for a library that they have just begun to use. 
Therefore, the ability to automatically identify idioms is needed.

In this paper, we present the first method for automatically mining code idioms
from an existing corpus of idiomatic code.
At first, this might seem to be a simple proposition: simply search for subtrees that occur often in a syntactically parsed corpus.  However, this naive method does not
work well, for the simple reason that frequent trees are not necessarily interesting trees. To return to our previous example, \lstinline{for} loops are much more common
than \lstinline{for} loops that iterate over arrays, but one would be hard pressed
to argue that \lstinline|for(...) {...}| on its own (that is, with no expressions
or body) is an interesting pattern.  

Instead, we rely on a different principle: interesting patterns are those 
that help to explain the code that programmers write.  As a measure of ``explanation quality'', we use a probabilistic model of the source code, and retain those idioms
that make the training corpus more likely under the model.  
These ideas can be formalized in a single, theoretically principled framework 
using a \emph{nonparametric Bayesian} analysis.
Nonparametric Bayesian methods have become enormously popular in statistics,
machine learning, and natural language processing because they provide
a flexible and principled way of automatically inducing a ``sweet spot'' of model
complexity based on the amount of data that is available \cite{orbanz10,gershman2012tutorial,TehJor2010a}.
In particular, we employ a \emph{nonparametric Bayesian tree substitution grammar},
which has recently been developed for natural language
\cite{cohn2010inducing,post2009bayesian}, but which has not been applied to source code.

Because our method is primarily statistical in nature, it is language agnostic,
and can be applied to any programming language for which one can collect a corpus
of previously-written idiomatic code.
Our major contributions are:

\begin{compactitem}
  \item We introduce the idiom mining problem (\autoref{sec:definition});
  
  \item We present \haggis, a method for automatically mining code idioms
  based on nonparametric Bayesian tree substitution grammars (\autoref{sec:tsg});
  
  \item We demonstrate that \haggis successfully identifies cross-project
  idioms (\autoref{sec:evaluation}), for example,  $67\%$ of idioms that we identify from one set of open source
  projects also appear in an independent set of snippets of example code from the popular Q\&A site StackOverflow;
  
  \item Examining the most common idioms that \haggis identifies (\autoref{fig:snippets}), we find that they describe 
  important program concepts, including object creation,
  exception handling, and resource management; 
  
  \item To further demonstrate that the idioms identified by \haggis
  are semantically meaningful, we examine the relationship between idioms
  and code libraries (\autoref{sec:libraries}), finding that many idioms 
  are strongly connected to package imports in a way that can support 
  suggestion.
\end{compactitem}

\section{Problem Definition}
\label{sec:definition}\label{sec:simple}

\begin{figure*}
\begin{center}
	\scriptsize
\begin{tabular}{c@{\hspace{3em}}cc}
\begin{minipage}{0.3\textwidth}
\begin{lstlisting}
...
if (c != null) {
  try {
   if (c.moveToFirst()) {
     number = c.getString(
               c.getColumnIndex(
                phoneColumn));
   }
  } finally {
   c.close();
  }
}
...
\end{lstlisting}
\end{minipage} 
&
\multirow{3}{*}[8em]{
\begin{minipage}{0.4\textwidth}
	\DTsetlength{0.2em}{0.4em}{0.2em}{0.4pt}{1.6pt}
	\scriptsize
	\dirtree{%
.1 IfStatement.
.2 \emph{expression:}.
.3 c!=null.
.2 \emph{then:}Block.
.3 \colorbox{light-gray}{TryStatement}.
.4 \colorbox{light-gray}{\emph{body:}IfStatement}.
.5 \colorbox{light-gray}{\emph{expr:}MethodInvocation}.
.6 \colorbox{light-gray}{\emph{expr:}var\%android.database.Cursor\%}.
.7 \emph{name:}c.
.6 \colorbox{light-gray}{\emph{name:}moveToFirst}.
.5 \colorbox{light-gray}{\emph{then:}Block}.
.6 number = c.getString(c.getColumnIndex(phoneColumn));.
.4 \colorbox{light-gray}{\emph{finally:}Block}.
.5 \colorbox{light-gray}{ExpressionStatement}.
.6 \colorbox{light-gray}{MethodInvocation}.
.7 \colorbox{light-gray}{\emph{expr:}var\%android.database.Cursor\%}.
.8 \emph{name:}c.
.7 \colorbox{light-gray}{\emph{name:}close}.
}
\end{minipage}}
&
\begin{minipage}{0.2\textwidth}
\hspace*{-2em}\begin{center}
$$E \rightarrow
\Tree [.$E$ [.$T$ $F$ * [.$F$ ( [.$E$ $T$ + $T$ ] !\qsetw{1cm} ) ] !\qsetw{1cm} ] ]$$\\ (prob $0.5$) \\[2ex]
(d)
\end{center}
\end{minipage}
\\
(a) \vspace{1em}& \\
\begin{minipage}{0.3\textwidth}
\begin{lstlisting}
try {
  if ($(Cursor).moveToFirst()) {
     $BODY$
  }
} finally {
   $(Cursor).close();
}
\end{lstlisting}
\end{minipage}
\\
(b) & (c) 
\end{tabular}
\end{center}

\caption{An example of how code idioms are extracted from ASTs.
(a) A snippet of code from the \texttt{PhoneNumberUtils} in 
the GitHub project \texttt{android.telephony}. (b) A commonly occurring idiom
when handling \texttt{android.database.Cursor} objects. This idiom is
successfully discovered by \haggis. (c) A partial representation
of the AST returned by the Eclipse JDT for the code in (a). 
The shaded nodes are those that are included in the idiom. 
(d) An example of a pTSG rule for a simple expression grammar.
See text for more details.}
\label{fig:motivational}
\end{figure*}

A \emph{code idiom} is a syntactic fragment that recurs across
software projects and serves a single semantic purpose.
An example of an idiom is shown in \autoref{fig:motivational}(b).
This is an idiom
which is used for manipulating objects of type \texttt{android.database.Cursor},
which ensures that the cursor is closed after use. (This idiom is indeed
discovered by our method.)
As in this example, typically idioms have parameters, which we will call \emph{metavariables}, such as the name of the \texttt{Cursor} variable,
and a code block describing what should be done if the \texttt{moveToFirst}
operation is successful.  An Android programmer who is unfamiliar
with this idiom might make bad mistakes, like not calling the \texttt{close} method
or not using a \lstinline{finally} block, causing subtle
memory leaks.

Many idioms, like the \texttt{close} example or those in 
\autoref{fig:snippets}, are specific to particular software libraries.
Other idioms are general across projects of the same programming language, such as those
in \autoref{fig:sampleIdioms}, including an idiom for looping over
an array or an idiom defining a \lstinline{String} constant.
(Again, all of the idioms in these figures are discovered
automatically by our method.)
Idioms concerning exception handling and resource management
are especially important to identify and suggest, because
failure to use them correctly can cause the
software to violate correctness properties.
As these examples show, idioms are usually \emph{parameterized}
and the  parameters often have syntactic
structure, such as expressions and code blocks.

We define idioms formally as fragments of abstract syntax
trees, which allows us to naturally represent the syntactic structure
of an idiom. More formally, an idiom is a fragment $\calT = (V,E)$
of an abstract syntax tree (AST), by which we mean the following.  Let $G$ be the context-free grammar of the programming language in question.  Then a fragment $\calT$ is a tree of
terminals and nonterminal from $G$ that is
a subgraph of some valid parse tree from $G$.\footnote{As a technicality, programming language grammars typically
describe parse trees rather than AST, but as there is a 1:1 mapping between the two,
we will assume that we have available a CFG that describes ASTs directly.}

An idiom $\calT$ can have as leaves both terminals and non-terminals.
Non-terminals
correspond to metavariables which must be filled in when instantiating the idiom.
For example, in \autoref{fig:motivational}(c), the shaded lines represent
the fragment for an example idiom; notice how the \texttt{Block}
node of the AST, which is a non-terminal, corresponds to a \lstinline{$BODY$}
metavariable in the pattern. 

\boldpara{Idiom mining}
Current IDEs provide tools for manually defining idioms and inserting
them when required, but this requires that the developer incur the
required setup cost, and that the developer know the idioms in the first place.
To eliminate these difficulties, we introduce the \emph{idiom mining problem}, namely, to identify a set of idioms
automatically given only a corpus of previously-written idiomatic code.
More formally, given a training set of source files with abstract syntax trees
 $\Data = \{ T_1, T_2, \ldots T_N \}$, the idiom mining problem is 
to identify a set of idioms $\Id = \{ \calT_i \}$
that occur in the training set.
This is an \emph{unsupervised} learning problem, as we do not assume
that we are provided with any example idioms that are explicitly identified.
Each fragment $\calT_i$ should occur as a subgraph
of every tree in some subset $\Data(\calT_i) \subseteq \Data$ of the training corpus.

\boldpara{What Idioms are Not}
Idioms are not clones.
A large amount of work in software engineering considers the
problem of clone detection \cite{roy2007survey,roy2009comparison}, some of which considers
\emph{syntactic} clones \cite{baker1993program,jiang2007deckard,kontogiannis1996pattern},
which find clones based on information from the AST.  Clones are contiguous blocks of code
that are used verbatim (or nearly so) in different code locations,
usually within a project and often created via copy-paste operations.
Idioms, on the other hand, typically recur \emph{across} projects,
even those from very different domains, and are used independently
by many different programmers.  Additionally, idioms are typically
not contiguous; instead, they have metavariables that bind to expressions or entire
code blocks.  Finally, idioms have a semantic purpose that developers are consciously
aware of. Indeed, we hypothesize that programmers chunk idioms into single mental units,
and often type them in to programs directly by hand, although
the psychological research necessary to verify this conjecture
is beyond the scope of the current paper.

Also, idiom mining is not API mining.  API mining \cite{nguyen2009graph,wang2013mining,zhong2009mapo}
is an active research area that focuses on mining groups of library functions
from the same API that are commonly used together.  These types of patterns
that are inferred are essentially sequences, or sometimes finite state machines,
of method invocations. Although API patterns have the potential to be 
extremely valuable to developers, idiom mining is a markedly different
problem because idioms have syntactic structure.
For example, current API mining approaches cannot find patterns 
 such as a library with a Tree class that requires special iteration
logic, or a Java library that requires the developer to free resources within a \lstinline|finally|
block.  These are exactly the type of patterns that \haggis identifies.

\boldpara{Simple Methods Do Not Work}
A natural first approach to this problem is  
to search for AST fragments that occur frequently, for example, to return the set of all fragments 
that occur more than a user-specified parameter $M$ times in the training set.  
This task is called frequent tree mining, and has been
the subject of some work in the data mining literature \cite{jimenez10frequent,termier02treefinder,zaki02efficiently,zaki05efficiently}.
Unfortunately, our preliminary investigations \cite{kuzborskij2011large}
found that these approaches do not yield good idioms.
Instead, the fragments that are returned tend to be small and generic,
omitting many details that, to a human eye, are central to the idiom.
For example, given the idiom in \autoref{fig:motivational}(c), 
it would be typical for tree mining methods to return a fragment
containing the \lstinline|try|, \lstinline|if|, and \lstinline|finally|
nodes but not the crucial method call to \lstinline|Cursor.close()|.

The reason for this is simple: Given a fragment $\calT$ that represents a true idiom,
it can always be made more frequent by removing one of the leaves,
even if it is strongly correlated with the rest of the tree.
So tree mining algorithms will tend to return these shorter trees,
resulting in incomplete idioms.
In other words, \emph{frequent patterns can be boring patterns.}
To avoid this problem, we need a way of penalizing the method when
it chooses \emph{not} to extend a pattern to include a node that
co-occurs frequently.  This is exactly what is provided by our 
probabilistic  approach.

\section{Mining Code Idioms}
\label{sec:tsg}

In this section, we introduce the technical framework that is required for
\haggis,\footnote{Holistic, Automatic Gathering of Grammatical Idioms from Software.} our proposed method for the idiom mining problem.
At a high level,
we approach the problem of mining source code idioms as that of 
inferring of commonly reoccurring fragments in ASTs. 
But, as we have seen, simple methods of formalizing this intuition do not work (see \autoref{sec:simple}), 
we resort to methods that are not as simple.  We apply recent advanced
techniques from statistical NLP \cite{cohn2010inducing,post2009bayesian},
but we need to explain them in some
detail to justify why they are appropriate for this software engineering
tasks, and why technically simpler methods would not be effective.

We will build up step by step.  First, we will describe two \emph{syntactic} probabilistic models
of source code, probabilistic context-free grammars and probabilistic tree substitution
grammars (pTSG). We will explain why pTSGs provide a straightforward framework
for augmenting a simple CFG to represent idioms.  The reason that we employ
\emph{probabilistic} models here, rather than a standard deterministic CFG,
is that probabilities provide a natural quantitative measure of the quality
of a proposed idiom: A proposed idiom is worthwhile only if, when we include
it into a pTSG, it increases the probability that the pTSG assigns to the training
corpus.

At first, it may seem odd that we apply grammar learning methods here, 
when of course the grammar of the programming language is already known. 
We clarify that our aim is \emph{not} to re-learn the known grammar, but rather
to learn probability distributions over parse trees from the known grammar.
These distributions will represent which rules from the grammar are used more often,
and, crucially, which rules tend to be used contiguously.

The pTSG provides us with a way to \emph{represent} idioms, but then we still need a way to \emph{discover} them.
It is for this purpose that we employ nonparametric Bayesian methods, 
a powerful general framework that provides methods that
automatically infer from data how complex a model should be.
After describing nonparametric Bayesian methods, we will finally
describe how to apply nonparametric Bayesian methods to pTSGs,
which requires a particular approximation known as Markov chain Monte Carlo.

\subsection{Probabilistic Grammars}

A \emph{probabilistic context free grammar} (PCFG) is 
a simple way to define a distribution over the strings of 
a context-free language. A PCFG is defined as $G = (\Sigma,N,S,R,\Pi)$,
where $\Sigma$ is a set of terminal symbols, $N$ a set of nonterminals, $S \in N$ is the root nonterminal
symbol and $R$ is a set of productions.  Each production in $R$ has the form $X \rightarrow Y$, where $X \in N$ and
$Y \in (\Sigma \cup N)^*$. The set $\Pi$ is a set of distributions $P(r|c),$
where $c \in N$ is a non-terminal, and $r \in R$ is a rule with $c$ on its left-hand side. 
To sample a tree from a PCFG, we recursively expand the tree, beginning
at $S$, and each time we add a non-terminal $c$ to the tree, we expand $c$ using
a production $r$ that is sampled from the corresponding distribution $P(r|c)$.
The probability of generating a particular tree $T$ from this procedure is 
simply the product over all rules that are required to generate $T$.  
The probability $P(x)$ of a string $x \in \Sigma^*$ is the sum of the probabilities 
of the trees $T$ that yield $x$, that is, we simply consider $P(x)$ as a
marginal distribution of $P(T)$.

\boldpara{Tree Substitution Grammars}
A tree substitution grammar (TSG) is a simple extension to a CFG,
in which productions expand into tree fragments
rather than simply into a list of symbols.
Formally, a TSG is also a tuple $G = (\Sigma,N,S,R)$, where
$\Sigma,N,S$ are exactly as in a CFG, but now each production
$r \in R$ takes the form $X \rightarrow \calT_X$, where $\calT_X$ is a fragment.
To produce a string from a TSG, we begin with a tree containing only $S$,
and recursively expanding the tree in a manner 
exactly analogous to a CFG --- the only difference
is that some rules can increase the height of the tree by more than 1.
A probabilistic tree substitution grammar (pTSG) $G$ \cite{cohn2010inducing,post2009bayesian} augments a TSG with probabilities,
in an analogous way to a PCFG. A pTSG is 
defined as $G=(\Sigma,N,S,R,\Pi)$ where $\Sigma$ is a set of terminal symbols,
$N$ a set of non terminal symbols, $S \in N$ is the root non-terminal
symbol, $R$ is a set of tree fragment productions. Finally, $\Pi$ is a set of distributions 
 $P_{TSG}(\calT_X|X)$, for all $X \in N$, each of which is a distribution over the set of all
 rules $X \rightarrow \calT_X$ in $R$ that have left-hand side $X$.
 
The key reason that we use pTSGs for idiom mining is that 
each tree fragment $\calT_X$ can be thought of as describing a set
of context-free rules that are typically used in sequence.
This is exactly what we are trying to discover in the idiom mining problem.
In other words, \emph{our goal will be to induce a pTSG in which every tree fragment
represents a code idiom} if the fragment has depth greater than $1$,
or a rule from the language's original grammar if the depth equals $1$.
As a simple example, consider the PCFG
\begin{alignat*}{5}
E &\rightarrow E + E  &\quad & \text{(prob $0.7$)}
  &\qquad\quad T &\rightarrow F * F  \quad &&\text{(prob $0.6$)} \\
E &\rightarrow T  &\quad & \text{(prob $0.3$)}
  &\qquad\quad T &\rightarrow F  \quad &&\text{(prob $0.4$)} \\
F &\rightarrow (E)   &\quad & \text{(prob $0.1$)}
  &\qquad\quad F &\rightarrow id  \quad &&\text{(prob $0.9$)},
\end{alignat*}
where $x$ and $y$ are non-terminals, and $E$ the start symbol.
Now, suppose that we are presented with a corpus of strings
from this language that include many instances of expressions
like $id * (id + id)$ and $id * (id + (id + id))$
(perhaps generated by a group of students who are practicing
the distributive law).  Then, we might choose to add a single pTSG rule to this
grammar, displayed in \autoref{fig:motivational}(d),
adjusting the probabilities for that rule and the
$E \rightarrow T + T$ and $E \rightarrow T$ rules
so that the three probabilities sum to 1.
Essentially, this allows us to a represent a
correlation between the rules $E \rightarrow T + T$
and $T \rightarrow F * F$. 

Finally, note that every CFG can be written
as a TSG where all productions expand to trees of depth 1.
Conversely, every TSG can be converted into
an equivalent CFG by adding extra non-terminals
(one for each TSG rule $X \rightarrow \calT_X$).
So TSGs are, in some sense, fancy notation for CFGs.
This notation will prove very useful, however, when we describe the learning
problem next.

\boldpara{Learning TSGs}
Now we define the learning problem for TSGs that we will consider.
First, we say that a pTSG $G_1 = (\Sigma_1, N_1, S_1, R_1, P_1)$ \emph{extends} a CFG $G_0$ if every tree with positive probability under $G_1$ is grammatically valid
according to $G_0$.  Given any set $\calT$ of tree fragments from $G_0$, we
can define a pTSG $G_1$ that extends $G_0$ as follows.
First, set $(\Sigma_1, N_1, S_1) = (\Sigma_0, N_0, S_0)$. Then, set $R_1 = R_{CFG} 
\cup R_{FRAG}$, where $R_{CFG}$ is the set of all rules from $R_0$, expressed in the 
TSG form, i.e., with right-hand sides as trees of depth 1, and 
$R_{FRAG}$ is a set of \emph{fragment rules} $X_i \rightarrow \calT_i$, for all $\calT_i \in \calT$ and where $X_i$ is the root of $\calT_i$.

The grammar learning problem that we consider can be called the \emph{\problem}.
The input is a set of trees $T_1 \ldots T_N$
from a context-free language with grammar $G_0 = (\Sigma_0, N_0, S_0, R_0)$.
The \problem is simply to learn a pTSG $G_1$ that
extends $G_0$ and is good at explaining the training set $T_1 \ldots T_N$.
The notion of ``good'' is deliberately vague; formalizing it is part of the problem. 
It should also be clear that we \emph{are not}
trying to learn the CFG for the original programming language --- instead,
we are trying to identify sequences of rules from the known
grammar that commonly co-occur contiguously.  

A na\"ive idea is to use \emph{maximum likelihood},
that is, to find the pTSG $G_1$ that extends $G_0$ and maximizes the
probability that $G_1$ assigns to $T_1 \ldots T_N$.
This does not work. The reason is that a trivial solution
is simply to add a fragment rule $E \rightarrow T_i$ for every
training tree $T_i$. This will assign a probability of $1/N$ to each training
tree, which in practice
will usually be optimal.  What is going on here is that the maximum likelihood
grammar is overfitting.  It is not surprising that this happens: 
there are an infinite number of potential trees that could be used to extend $G_0$,
so if a model is given such a large amount of flexibility, overfitting becomes inevitable.
What we need is a strong method of controlling overfitting, which 
the next section provides.

%

\subsection{Nonparametric Bayesian Methods}

At the heart of any application of machine learning is the need 
to control the complexity of the model. For example,
in a clustering task, 
many standard clustering methods, such as $K$-means, require the user to pre-specify the number of clusters $K$ in advance.  If $K$ is too small, then each cluster
will be very large and not contain useful information about the data.
If $K$ is too large, then each cluster will only contain a few data points,
so the again, the cluster centroid will not tell us much about the data set.
For the \emph{\problem}, the key factor that determines model complexity is the number of
fragment rules that we allow for each non-terminal. 
If we allow the model to assign too many fragments to each non-terminal,
then it can simply memorize the training set, as described in the previous
section.
But if we allow too few, then the model will be unable to find
useful patterns.
Nonparametric Bayesian methods provide a powerful and
theoretically principled method for managing this trade-off.

To explain how this works, we must first
explain {Bayesian statistics}. Bayesian statistics
\cite{gelman:bda,murphy2012machine} is alternative general framework to classical {frequentist}
statistical methods
such as confidence intervals and hypothesis testing.  The idea behind
Bayesian statistics is that whenever one wants to estimate an unknown
quantity $\theta$ from a data set $x_1, x_2, \ldots x_N$, the analyst should
choose a prior distribution $P(\theta)$ that encodes any prior knowledge
about $\theta$
(if little is known, this distribution can be vague), 
and then a model $P(x_1 \ldots x_N \,|\, \theta)$.  Once we define
these two quantities, the laws of probability provide only one choice 
for how to infer $\theta$, which is to compute the conditional distribution
$P(\theta | x_1 \ldots x_N)$ using Bayes' rule.  This distribution
is called the \emph{posterior distribution} and encapsulates
all of the information that we have about $\theta$ from the data.
Bayesian methods provide powerful general tools to combat overfitting, 
as the prior $P(\theta)$ can be chosen to encourage simpler models.

If $\theta$ is a finite-dimensional set of parameters, such as the mean and the
variance of a Gaussian distribution, then it is easy to construct
an appropriate prior $P(\theta)$.
Constructing a prior becomes more difficult, however, when $\theta$ does not
have a fixed number of dimensions, which is what occurs when we wish to infer
the model complexity automatically.  For example, consider a clustering
model, where we want to learn the number of clusters.  In this case, $\theta$ would be a vector
containing the centroid for each cluster, but then, because before we see the data
the number of clusters could be arbitrarily large, $\theta$ has unbounded dimension.  As another example, in the case of the \problem, we do not know
in advance how many fragments are associated with each non-terminal,
and so want to infer this from data.
\emph{Nonparametric Bayesian methods} focus on developing prior distributions
over infinite dimensional objects, which are then used within Bayesian statistical
inference. Bayesian nonparametrics  have been the subject of intense research 
in statistics and in machine learning, with popular models including 
the Dirichlet process \cite{hjort2010bayesian} and the Gaussian process \cite{williams2006gaussian}.

Applying this discussion to the \problem, what we are trying to infer
is a pTSG, so, to apply Bayesian inference, our prior distribution must be \emph{a probability distribution over probabilistic grammars}.  
We will bootstrap this from a distribution over context-free fragments, which we define first.
Let $G_0$ be the known context-free grammar for the programming language in question.
We will assume that we have available a PCFG for $G_0$, because this can be easily estimated by maximum likelihood from our training corpus; call this distribution
over trees $P_{\ML}$.  Now, $P_{\ML}$ gives us a distribution over full trees,
but what we will require is a distribution over \emph{fragments}.
We define this simply as
\begin{align}
	P_{0}(T)=P_{\geom}\left(|T|,p_{\$}\right)\prod_{r \in T}P_{\ML}(r),
\end{align}
where $|T|$ is the size of the fragment $T$, $P_{\geom}$ is a geometric
distribution with parameter $p_{\$}$, and $r$ ranges over the multiset
of productions that are used within $T$.

Now we can define a prior distribution over pTSGs.
Recall that we can define a pTSG $G_1$ that extends $G_0$ by
specifying a set of tree fragments $\setF_X$ for each non-terminal $X$.
So, to define a distribution over pTSGs, we will define a distribution
$P(\setF_X)$ over the set of tree fragments rooted at $X$.
We need $P(\setF_X)$ to have several important properties. 
First, we need $P(\setF_X)$ to have infinite support, that is,
it must assign positive probability to \emph{all possible fragments}.
This is because if we do not assign a fragment positive probability
in the prior distribution, 
we will never be able to infer it as an idiom, no matter how often it appears.
Second, we want $P(\setF_X)$ to exhibit a ``rich-get-richer'' effect,
namely, once we have observed that a fragment $\calT_X$ occurs many times,
we want to be able predict that it will occur more often in the future.

The simplest distribution that has these properties is the Dirichlet process (DP).
The Dirichlet process has two parameters: a \emph{base measure},\footnote{The
base measure will be a probability measure, so for our purposes here,
we can think of this as a fancy word for ``base distribution''.}
which in our case will be the fragment distribution $P_0$, and a concentration parameter $\alpha \in \R^+$, which controls how strong the rich-get-richer effect is.  
One simple way to characterize the Dirichlet process 
is the \emph{stick-breaking} representation \cite{sethuraman1991constructive}.
Using this representation,
a Dirichlet process defines a distribution over $\setF_X$ as
\begin{alignat*}{2}
\Prob{\calT \in \setF_X}&= \sum_{k=1}^{\infty} \pi_k \delta_{\{\calT = \calT_k\}} 
\qquad\quad &\calT_k &\sim P_0 \\[-2ex]
u_k &\sim \mbox{Beta}(1,\alpha)  
&\pi_k &= (1-u_k)\prod_{j=1}^{k-1} u_j.  
\end{alignat*}
To interpret this, recall that the symbol $\sim$ is read ``is distributed as,''
and the Beta distribution is a standard distribution over the set $\lbrack 0, 1 \rbrack$; as $\alpha$ becomes large, the mean of a Beta$(1,\alpha)$ distribution
will approach $1$.  Intuitively, what is going on here is that a sample from 
the DP is a distribution over a countably infinite number of fragments
$\calT_1, \calT_2, \ldots$. Each one of these fragments is sampled 
independently from the fragment distribution $P_0$.  To assign a probability to each
fragment, we recursively split the interval $[0,1]$ into a countable number of sticks
$\pi_1, \pi_2, \ldots$. The value $(1-u_k)$ defines what proportion of the remaining
stick is assigned to the current sample $\calT_k$, and the remainder
is assigned to the infinite number of remaining trees $\calT_{k+1}, \calT_{k+2}, \ldots$.  This process defines a distribution over fragments
$\setF_X$ for each non-terminal $X$, and hence a distribution $P(G_1)$ over
the set of all pTSGs that extend $G_0$.  We will refer to this distribution
as a \emph{Dirichlet process probabilistic tree substitution grammar} (DPpTSG)
\cite{post2009bayesian,cohn2010inducing}.

This process may seem odd for two reasons:
(a) each sample from $P(G_1)$ is infinitely large, so we cannot store
it exactly on a computer, (b) the fragments from $G_1$ are sampled randomly
from a PCFG, so there is no reason to think that they should match real idioms.
Fortunately, the answer to both these concerns is simple. We are \emph{not} interested in the fragments
that exist in the prior distribution, but rather of those in the posterior
distribution.  More formally, the DP provides us with a prior distribution $G_1$
over pTSGs. But $G_1$ itself, like any pTSG,
defines a distribution $P(T_1, T_2, \ldots T_N | G_1)$ over the training set.
So, just as in the parametric case, we can apply Bayes's rule to obtain
a posterior distribution $P(G_1 | T_1, T_2, \ldots T_N)$. It can be
shown that this distribution is also a DPpTSG, and, amazingly,
that this posterior DPpTSG can be characterized by a \emph{finite}
set of fragments $\setF'_X$ for each non-terminal.  It is these fragments
that we will identify
as code idioms (\autoref{sec:tsgForCode}).

\subsection{Inference}

\begin{figure}
	\centering
	\begin{tikzpicture}[node distance=1.5cm,>=stealth',bend angle=45,auto,scale=0.7, every node/.style={scale=0.7}]
	  \tikzstyle{place}=[circle,thick,draw=blue!75,fill=blue!20,minimum size=6mm]
	  \tikzstyle{terminal}=[circle,thick,double distance=0.2mm,draw=blue!75,fill=blue!20,minimum size=6mm]
	  \tikzstyle{transition}=[rectangle,thick,draw=black!75,
				  fill=black!20,minimum size=4mm]
	  \tikzstyle{dots}=[rectangle,thin,
				  minimum size=0.1mm]

	  \begin{scope}
		\node [place,tokens=1]     (for)     {};
			\node [place]     (vardecl) [below left of=for] {} edge [pre] node {}(for);
				\node [terminal]     (vardecltype) [below left of=vardecl] {} edge [pre] node {}(vardecl);
				\node [place, tokens=1]     (vardeclfrag) [right of=vardecltype] {} edge [pre] node {}(vardecl);
					\node [place]     (vardeclfragC) [below left of=vardeclfrag] {} edge [pre] node {}(vardeclfrag);
						\node [dots]     (vardeclfragDots1) [below left of=vardeclfragC] {} edge [thick,dotted] node {}(vardeclfragC);
						\node [dots]     (vardeclfragDots2) [below of=vardeclfragC] {} edge [thick,dotted] node {}(vardeclfragC);

			\node [place]     (forexp) [right of=vardecl] {} edge [pre] node {}(for);
				\node [place]     (forexp1) [below of=forexp] {} edge [pre] node {}(forexp);
					\node [terminal]     (forexp2) [below left of=forexp1] {} edge [pre] node {}(forexp1);
				
			\node [place]     (forupd) [right of=forexp] {$s$} edge [pre] node {}(for);
				\node [place]     (Infix) [below of=forupd] {} edge [pre] node {}(forupd);
					\node [place]     (InfixNamTmp) [below left of=Infix] {} edge [pre] node {}(Infix);
						\node [terminal]     (InfixNam) [below left of=InfixNamTmp] {} edge [pre] node {}(InfixNamTmp);
				\node [place]     (Infix2) [right of=Infix] {} edge [pre] node {}(forupd);
					\node [place]     (Infix2exp) [below left of=Infix2] {} edge [pre] node {}(Infix2);					
						\node [terminal]     (Infix21) [below right of=Infix2exp] {} edge [pre] node {}(Infix2exp);
						\node [terminal]     (Infix22) [left of=Infix21] {} edge [pre] node {}(Infix2exp);
					\node [place,tokens=1]     (Infix2nam) [right of=Infix2exp] {} edge [pre] node {}(Infix2);
						\node [dots]     (Infix2namDots1) [below right of=Infix2nam] {} edge [thick,dotted] node {}(Infix2nam);				
						\node [dots]     (Infix2namDots2) [right of=Infix2namDots1] {} edge [thick,dotted] node {}(Infix2nam);	
					\node [terminal]     (Infix2othe) [right of=Infix2nam] {} edge [pre] node {}(Infix2);
			
			\node [place,tokens=1]     (body) [right of=forupd] {} edge [pre] node {}(for);
				\node [dots]     (bodyD1) [below right of=body] {} edge [thick,dotted] node {}(body);
				\node [dots]     (bodyD1) [right of=bodyD1] {} edge [thick,dotted] node {}(body);
		
		\node [dots]     (t1annotation) [above left of=vardecl] {$\mathcal{T}_{t}$};
		\node [dots]     (t2annotation) [right of=Infix2othe,xshift=-0.8cm] {$\mathcal{T}_{s}$};		
		
	  \end{scope}

	  \begin{pgfonlayer}{background}
	\draw [fill=black!15,opacity=.5] (vardecltype.south -| vardecltype.west) 
		to [out=130,in=80] (vardecltype.north -| vardecltype.west)   
		to [out=80,in=190] (for.north -| for.west)  
		to [out=190,in=160] (for.north -| for.east)  
		to [out=0,in=160] (body.north -| body.east) 
		to [out=-20,in=5] (body.south -| body.east) 
		to [out=200,in=10] (forupd.south -| forupd.east) 
		to [out=200,in=30] (forexp2.south -| forexp2.east) 
		to [out=200,in=-30] (forexp2.south -| forexp2.west) 
		to [out=150,in=-30] (vardeclfrag.south -| vardeclfrag.west) 
		to [out=150,in=-70] (vardecltype.south -| vardecltype.west);
	
	\draw [fill=yellow!15,opacity=.5] (forupd.north -| forupd.east) 
		to [out=120,in=30] (forupd.north -| forupd.west)  
		to [out=200,in=40] (Infix.north -| forupd.west)  
		to [out=210,in=40] (InfixNam.north -| InfixNam.west)  
		to [out=220,in=160] (InfixNam.south -| InfixNam.west)  
		to [out=-20,in=200] (Infix21.south -| Infix21.east)  
		to [out=20,in=-20] (Infix2nam.south -| Infix2nam.west)  
		to [out=-0,in=-160] (Infix2othe.south -| Infix2othe.east)
		to [out=40,in=90] (Infix2othe.north -| Infix2othe.east)
		to [out=160,in=-50] (forupd.north -| forupd.east) 
		;		
	 \end{pgfonlayer}
	\end{tikzpicture}

\caption{Sampling an AST. Nodes with dots show the points where the tree
is split (\ie $z_t=1$). Nodes with double border represent terminal nodes.}
\label{fig:tsgSampling}
\end{figure}

What we have just discussed is how to define a posterior distribution over grammars 
that will infer code idioms. But we still need to describe how to \emph{compute}
this distribution. Unfortunately, the posterior
distribution cannot be computed exactly, so we resort to approximations.
The most commonly used approximations in the literature are based
on Markov chain Monte Carlo (MCMC), which we explain below.
But first, we make one more observation about pTSGs.
All of the pTSGs that we consider are extensions of an unambiguous
base CFG $G_0$.  This means that given a source file $F$, 
we can separate the pTSG parsing task
into two steps: first, parse $F$ using $G_0$, resulting in a CFG tree $T$; 
second, group the nodes in $T$ according to which fragment rule
in the pTSG was used to generated them.
We can represent this second task as a tree of binary variables $z_s$
for each node $s$. These variables indicate whether the node $s$ is the root
of a new fragment ($z_s = 1$), or if node $s$ is part of the same fragment
as its parent ($z_s = 0$). Essentially, the variables $z_s$ show 
the boundaries of the inferred tree patterns; see \autoref{fig:tsgSampling}
for an example.  Conversely, even if we don't know what fragments are
in the grammar, given a training corpus that has been parsed in this way,
we can use the $z_s$ variables to read off what fragments must
have been in the pTSG.

With this representation in hand, we are now 
ready to present an MCMC method for sampling from
the posterior distribution over grammars, using a particular
method called Gibbs sampling.
Gibbs sampling is an iterative method, which starts with an initial value
for all of the $z$ variables, and then updates them one at a time.
At each iteration, the sampler visits 
every tree node $t$ of every tree in the training corpus,
and samples a new value for $z_t$.  Let $s$ be the parent of $t$.
If we choose $z_t = 1$, we can examine the current values of the $z$ variables to determine
the tree fragment $\calT_t$ that contains $t$ and the fragment $\calT_s$
for $s$, which must be disjoint. On the other hand, if we set $z_t = 0$,
then $s$ and $t$ will belong to the same fragment, which
will be exactly $\calT_{\join} = \calT_s \cup \calT_t$.
Now, we set $z_t$ to $0$ with probability
\begin{align}
	P_{}(z_t=0) = \frac{P_{\post}(\mathcal{T}_{\join})}{P_{\post}(\mathcal{T}_{\join})+P_{\post}(\mathcal{T}_{s})P_{\post}(\mathcal{T}_{t})}.
\end{align}
where
\begin{align}
	P_{\post}(\calT) = \frac{\textsf{count}(\calT)+\alpha P_{0}(\calT)}{\textsf{count}(h(\calT))+\alpha},
\end{align}
$h$ returns the root of the fragment, and \textsf{count}
returns the number of times that a tree occurs as a fragment
in the corpus, as determined by the current values of $z$.
Intuitively, what is happening here is that if the fragments $\calT_s$
and $\calT_t$ occur very often together in the corpus, relative to the
number of times that they occur independently, then we are more likely
to join them into a single fragment.

It can be shown that if we repeat this process for a large number
of iterations, eventually the resulting distribution over fragments
will converge to the posterior distribution over fragments defined
by the DPpTSG.
It is these fragments that we return as idioms.

We present the Gibbs sampler because it is a useful illustration
of MCMC, but in practice we find that it converges
too slowly to scale to large codebases.
Instead we use the type-based MCMC 
sampler of Liang \etal \cite{liang2010type} (details omitted).

\section{Sampling a TSG for Code}
\label{sec:tsgForCode}
Hindle \etal \cite{hindle2012naturalness} have shown that source code
presents some of the characteristics of natural language. \haggis exploits this
fact by using pTSGs --- originally devised for natural language --- to
infer code idioms. Here, we describe a set of necessary transformations
to ASTs and pTSG to adapt these general methods specifically to the task
of inferring code idioms.

\boldpara{AST Transformation}
For each \texttt{.java} file we use the Eclipse JDT \cite{eclipseJdt}
to extract its AST --- a tree structure of \texttt{ASTNode} objects.
Each \texttt{ASTNode} object contains two sets of properties: 
\emph{simple properties} --- such as the type of the operator, if 
\texttt{ASTNode} is an infix expression --- and \emph{structural properties}
that contain zero or more child \texttt{ASTNode} objects. First, we
construct the grammar symbols by mapping each \texttt{ASTNode}'s type 
and simple properties into a (terminal or non-terminal) symbol.
The transformed tree is then constructed by mapping the original AST
into a tree whose nodes are annotated with the symbols. Each node's 
children are grouped by property.

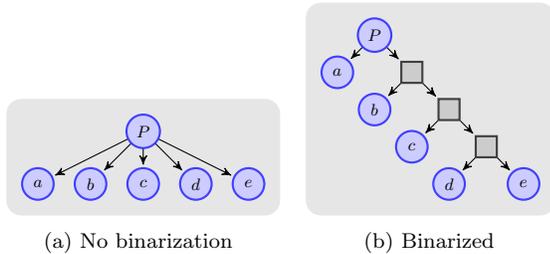
\begin{figure}
	\centering

\begin{subfigure}[b]{0.2\textwidth}
	\centering
	\begin{tikzpicture}[node distance=1cm,>=stealth',bend angle=45,scale=0.7, every node/.style={scale=0.7}]

	  \tikzstyle{place}=[circle,thick,draw=blue!75,fill=blue!20,minimum size=6mm]
	  \tikzstyle{transition}=[rectangle,thick,draw=black!75,
				  fill=black!20,minimum size=4mm]

	  \begin{scope}
		\node [place] (p1)                          {$P$};    
		\node [place] (c3) [below of=p1]   {$c$} edge [pre](p1) ;
		\node [place] (c2) [right of=c3]  {$d$} edge [pre](p1) ;
		\node [place] (c1) [right of=c2]  {$e$} edge [pre](p1) ;
		\node [place] (c4) [left of=c3]   {$b$} edge [pre](p1) ;
		\node [place] (c5) [left of=c4]   {$a$} edge [pre](p1) ;
	  \end{scope}

	  \begin{pgfonlayer}{background}
		\filldraw [line width=4mm,join=round,black!10]
		  (p1.north  -| c1.east)  rectangle (c5.south  -| c5.west);
	  \end{pgfonlayer}
	\end{tikzpicture}
	\caption{No binarization}
\end{subfigure}
~
\begin{subfigure}[b]{0.2\textwidth} \centering
	\begin{tikzpicture}[node distance=1cm,>=stealth',bend angle=45,auto,scale=0.7, every node/.style={scale=0.7}]
	  \tikzstyle{place}=[circle,thick,draw=blue!75,fill=blue!20,minimum size=6mm]
	  \tikzstyle{transition}=[rectangle,thick,draw=black!75,
				  fill=black!20,minimum size=4mm]

	  \begin{scope}[xshift=6cm]
		\node [place]     (p1')                                                {$P$};
		\node [place]     (c1') [below left of=p1'] {$a$} edge [pre](p1');    
		\node [transition] (e1') [below right of=p1'] {} edge [pre](p1');
		\node [place]     (c2') [below left of=e1'] {$b$} edge [pre](e1');
		\node [transition] (e2') [below right of=e1'] {} edge [pre](e1'); 
		\node [place]     (c3') [below left of=e2'] {$c$} edge [pre](e2');
		\node [transition] (e3') [below right of=e2'] {} edge [pre](e2'); 
		\node [place]     (c4') [below left of=e3'] {$d$} edge [pre](e3');
		\node [place]     (c5') [below right of=e3'] {$e$} edge [pre](e3');
	  \end{scope}

	  \begin{pgfonlayer}{background}
		\filldraw [line width=4mm,join=round,black!10]
		  (p1'.north -| c1'.west) rectangle (c5'.south -| c5'.east);
	  \end{pgfonlayer}
	\end{tikzpicture}
	\caption{Binarized}
\end{subfigure}
\caption{Tree Binarization for nodes with multiple children. Square 
nodes represent the dummy nodes added.}
\label{fig:binarization}
\end{figure}

The transformed trees may contain nodes that have more than two children for
a single property (\eg \texttt{Block}). This induces unnecessary sparsity
in the CFG and TSG rules. To reduce this sparsity, we perform \emph{tree binarization}.
This process --- common in NLP --- transforms
the original tree into binary by adding dummy nodes, making the data less
sparse. It will also help us capture idioms in sequential statements. 
Note that binarization is performed \emph{only} on structural properties
that have two or more children, while an arbitrary node may have
more than two children among its properties. 

One final hurdle for learning meaningful code idioms are variable
names. Since variable names are mostly project or class specific we
abstract them introducing an intermediate \texttt{MetaVariable} node
between the  \texttt{SimpleName} node containing the string representation of the 
variable name and its parent node. \texttt{MetaVariable} nodes are also annotated
with the type of the variable they are abstracting. This provides the pTSG 
with the flexibility to either exclude or include variable names as appropriate.
For example, in the snippet of
\autoref{fig:motivational}(a) by using metavariables, we are able to
learn the idiom in \autoref{fig:motivational}(b) without specifying the
name of the \lstinline{Cursor} object by excluding the \texttt{SimpleName} nodes
from the fragment. Alternatively, if
a specific variable name is common and idiomatic, such as the 
\lstinline{i} in a \lstinline{for} loop, the pTSG can choose to
include \texttt{SimpleName} in the extracted idiom, by merging it with its parent
\texttt{MetaVariable} node.

\boldpara{Training TSGs and Extracting Code Idioms}
Training a pTSG happens offline, during a separate training phase. After
training the pTSG, we then extract the mined code idioms which then can
be used for any later visualization. In other words,
a user of a \haggis IDE tool would never need to wait for a MCMC
method to finish.
The output of a MCMC method is a series of (approximate) samples from 
the posterior distribution, each of which in our case, is a single pTSG.
These sampled pTSGs need to be post-processed to extract a single, meaningful
set of code idioms.
First, we aggregate the MCMC samples after removing the first few samples as \emph{burn-in}, which is standard methodology for applying MCMC.
Then, to extract idioms from the remaining samples, we merge all samples'
tree fragments into a single multiset. We then prune the multiset by
removing all tree fragments that have been seen less than
$c_{min}$ times to ensure that the mined tree fragments are frequent enough.
We also prune fragments that have fewer that $n_{min}$ nodes to get
a set of non-trivial (\ie sufficiently large) code idioms.
Finally, we reconvert the fragments back to Java code. The leaf nodes of
the fragments that contain non-terminal symbols represent metavariables
and are converted to the appropriate symbol that is denoted by a \$ prefix.

Additionally, to assist the sampler in inducing meaningful idioms, we
prune any \lstinline{import} statements from the corpus, so that they 
cannot be mined as idioms. We also 
exclude some nodes from sampling, fixing $z_i=0$ and thus forcing some
nodes to be un-splittable. Such nodes include method invocation arguments,
qualified and parametrized type node children, non-block children of 
\lstinline{while}, \lstinline{for} and \lstinline{if} statement nodes,
parenthesized, postfix and infix expressions and variable declaration statements.

\section{Code Snippet Evaluation}
\label{sec:evaluation}
We take advantage of the omnipresence of idioms in source code
to evaluate \haggis on popular open source projects. We restrict ourselves to
the Java programming language, due to the high availability of tools
and source code.  We emphasize, however, that \haggis is language
agnostic.
Before we get started, an interesting way to get an intuitive feel
for any probabilistic model is simply to draw samples from it.
\autoref{fig:sample} shows a code snippet that we synthetically generated
by sampling from
the posterior distribution over code defined by the pTSG. One can observe that the pTSG
is learning to produce idiomatic and syntactically correct code, although
--- as expected --- the code is semantically inconsistent.

\begin{figure}[t]
\begin{center}
\footnotesize
\begin{lstlisting}[basicstyle=\scriptsize\ttfamily]
try {
	regions=computeProjections(owner);
} catch (RuntimeException e) {
	e.printStackTrace();
	throw e;
}
if (elem instanceof IParent) {
	IJavaElement[] children=((IParent)owner).getChildren();
	for (int fromPosition=0; i < children.length; i++) {
		IJavaElement aChild=children[i];
		Set childRegions=findAnnotations(aChild,result);
		removeCollisions(regions,childRegions);
	}
}
constructAnnotations(elem,result,regions);
\end{lstlisting}
\end{center}
\vspace{-2em}
\normalsize
\caption{Synthetic code randomly generated from  a posterior pTSG.
One can see that the pTSG produces code that is syntactically correct  
and locally consistent. This effect allows us to infer code idioms.
It can be seen that, as expected, the pTSG cannot capture higher level information,
such as variable binding.}
\label{fig:sample}

\end{figure}

\boldpara{Methodology}
We use two evaluation datasets comprised of Java open-source code
available on GitHub. The \projectsDset dataset (\autoref{tbl:inprojects}) contains the top
13 Java GitHub projects whose repository is at least 100MB in size 
according to the GitHub Archive \cite{githubarchive}.
To determine popularity, we computed the $z$-score of forks and watchers
for each project. The normalized scores were then averaged to retrieve each
project's popularity ranking. 
The second evaluation dataset, \libraryDset (\autoref{tbl:xprojects}),
consists of Java classes that import (\ie use) 15 popular Java libraries.
For each selected library, we retrieved from the Java GitHub Corpus
\cite{allamanis2013mining} all files that import that library but do not
implement it. We split both datasets into a train and a test set, splitting
each project in \projectsDset and each library fileset in \libraryDset into
a train (70\%) and a test (30\%) set. The \projectsDset will be used to
mine project-specific idioms, while the \libraryDset will be used to mine
idioms that occur across libraries.

\begin{figure}[tb]
\begin{center}
\scriptsize
\begin{tabular}{lrrrrp{2.5cm}} \toprule
Name & Forks & Stars&Files & Commit & Description \\ \midrule
\textsf{arduino} & 2633  & 1533 &180&\texttt{2757691}& Electronics Prototyping\\
\textsf{atmosphere} &  1606 & 370&328& \texttt{a0262bf}&WebSocket Framework\\
\textsf{bigbluebutton} &1018   & 1761& 760&\texttt{e3b6172}& Web Conferencing\\
\textsf{elasticsearch} & 5972 & 1534 &3525& \texttt{ad547eb} & REST Search Engine\\
\textsf{grails-core} &  936 & 492&831& \texttt{15f9114}& Web App Framework \\
\textsf{hadoop} &  756 & 742& 4985& \texttt{f68ca74}& Map-Reduce Framework\\
\textsf{hibernate} &  870 & 643& 6273& \texttt{d28447e}&ORM Framework\\
\textsf{libgdx} &  2903 &2342 &1985& \texttt{0c6a387}& Game Dev Framework\\
\textsf{netty} &  2639 & 1090 &1031& \texttt{3f53ba2} & Net App Framework\\
\textsf{storm} &  1534 & 7928 &448&\texttt{cdb116e}& Distributed Computation \\
\textsf{vert.x} &  2739 & 527 &383& \texttt{9f79416}&Application platform\\
\textsf{voldemort} &  347 & 1230 &936&\texttt{9ea2e95}& NoSQL Database\\
\textsf{wildfly} &  1060 & 1040 &8157&\texttt{043d7d5}&Application Server \\ \bottomrule
\end{tabular}
\end{center}
\normalsize
\caption{\projectsDset dataset used for in-project idiom evaluation. 
Projects in alphabetical order.}
\label{tbl:inprojects}
\end{figure}

\begin{figure}[tb]
\footnotesize
\begin{center}
	\scriptsize
\begin{tabular}{lrp{3cm}} \toprule
Package Name &Files & Description \\ \midrule
\texttt{android.location} &1262& Android location API\\
\texttt{android.net.wifi} &373& Android WiFi API\\
\texttt{com.rabbitmq} &242& Messaging system\\
\texttt{com.spatial4j} &65&Geospatial library\\
\texttt{io.netty} &65&Network app framework\\
\texttt{opennlp} &202&NLP tools\\
\texttt{org.apache.hadoop} &8467& Map-Reduce framework\\
\texttt{org.apache.lucene} &4595& Search Server\\
\texttt{org.elasticsearch} &338& REST Search Engine\\
\texttt{org.eclipse.jgit} &1350& Git implementation\\
\texttt{org.hibernate} &7822& Persistence framework\\
\texttt{org.jsoup} &335& HTML parser\\
\texttt{org.mozilla.javascript} &1002& JavaScript implementation\\
\texttt{org.neo4j} &1294& Graph database\\
\texttt{twitter4j} &454&Twitter API\\ \bottomrule
\end{tabular}
\end{center}
\normalsize
\caption{\libraryDset dataset for cross-project idiom evaluation.
Each API fileset contains all class files that \lstinline+import+ a class belonging
to the respective package or one of its subpackages.}
\label{tbl:xprojects}
\end{figure}

To extract idioms we run MCMC for 100 iterations for each of the projects in 
\projectsDset and each of library filesets in the \libraryDset
allowing sufficient burn-in time of 75 iterations. For the last 25 iterations,
we aggregate a sample posterior pTSG and extract idioms as detailed in
\autoref{sec:tsgForCode}.
A threat to the validity of the evaluation using the aforementioned datasets
is the possibility that the datasets are not representative of Java
development practices, containing solely open-source projects from
GitHub. However, the selected datasets span a wide variety of domains,
including databases, messaging systems and code parsers, diminishing any
such possibility. Furthermore, we perform an extrinsic evaluation on source
code found on a popular online Q\&A website, StackOverflow.

\boldpara{Evaluation Metrics}
We compute two metrics on the test corpora.
These two metrics resemble precision and recall in information retrieval
but are adjusted to the code idiom domain. We define \emph{idiom coverage}
as the percent of source code AST nodes that can be matched to the mined idioms.
Coverage is thus a number between 0 and 1 indicating the extent to which
the mined idioms exist in a piece of code. We define \emph{idiom set 
precision} as the percentage of the mined idioms found in the test corpus.
This metric shows the precision of mined set of idioms.
Using these two metrics, we also tune the concentration parameter
of the DPpTSG model by using \texttt{android.net.wifi} as a validation set,
yielding $\alpha=1$.

\begin{figure}[tb]
\centering
\begin{subfigure}[b]{.8\columnwidth}
	\centering
	\begin{lstlisting}[basicstyle=\scriptsize\ttfamily]
for (Iterator iter=$methodInvoc;
		iter.hasNext(); ) 
	{$BODY$}
\end{lstlisting}
\vspace{-1em}\caption{Iterate through the elements of an Iterator.}\vspace{0.25em}
\label{fig:iterator}
\end{subfigure}
~
\begin{subfigure}[b]{.8\columnwidth}
	\begin{lstlisting}[basicstyle=\scriptsize\ttfamily]
private final static Log $name=
	LogFactory.getLog($type.class);
\end{lstlisting}
\vspace{-1em}\caption{Creating a logger for a class.}\vspace{0.25em}
\label{fig:logger}
\end{subfigure}
~
\begin{subfigure}[b]{.8\columnwidth}
	\begin{lstlisting}[basicstyle=\scriptsize\ttfamily]
public static final 
	String $name = $StringLit;
\end{lstlisting}
\vspace{-1em}\caption{Defining a constant String.}\vspace{0.25em}
\label{fig:stringConst}
\end{subfigure}
~
\begin{subfigure}[b]{.8\columnwidth}
	\begin{lstlisting}[basicstyle=\scriptsize\ttfamily]
while (($(String) = $(BufferedReader).
		readLine()) != null) {
	$BODY$
}
\end{lstlisting}
\vspace{-1em}\caption{Looping through lines from a \texttt{BufferedReader}.}
\end{subfigure}
\caption{Sample Java-language idioms. \lstinline+$stringLit+ denotes a user-defined
string literal, \lstinline+$name+ a freely
defined (variable) name, \lstinline+$methodInvoc+ a single method
invocation statement, \lstinline+$ifstatement+ a single \lstinline+if+
statement and \lstinline+$BODY$+ denotes a user-defined
code block of one or more statements.}
\label{fig:sampleIdioms}
\end{figure}

\begin{figure*}[tbp]
\centering~
\begin{subfigure}[b]{0.3\textwidth}
\begin{lstlisting}[basicstyle=\footnotesize\ttfamily]
channel=connection.
	createChannel();
\end{lstlisting}
\caption{}
\label{fig:rabbitmqConnection}
\end{subfigure}
~
\begin{subfigure}[b]{0.3\textwidth}
\begin{lstlisting}[basicstyle=\footnotesize\ttfamily]
Elements $name=$(Element).
	select($StringLit);
\end{lstlisting}
\caption{}
\label{fig:jsoupSelect}
\end{subfigure}
~
\begin{subfigure}[b]{0.3\textwidth}
\begin{lstlisting}[basicstyle=\footnotesize\ttfamily]
Transaction tx=ConnectionFactory.
		getDatabase().beginTx();
\end{lstlisting}
\caption{}
\label{fig:neo4jTransaction}
\end{subfigure}
~
\begin{subfigure}[b]{0.3\textwidth}
\begin{lstlisting}[basicstyle=\footnotesize\ttfamily]
catch (Exception e) {
  $(Transaction).failure();
}
\end{lstlisting}
\caption{}
\label{fig:neo4jTransactionFailure}
\end{subfigure}
~
\begin{subfigure}[b]{0.3\textwidth}
\begin{lstlisting}[basicstyle=\footnotesize\ttfamily]
SearchSourceBuilder builder=
	getQueryTranslator().build(
		$(ContentIndexQuery));
\end{lstlisting}
\caption{}
\label{fig:elasticBuilder}
\end{subfigure}
~
\begin{subfigure}[b]{0.3\textwidth}
\begin{lstlisting}[basicstyle=\footnotesize\ttfamily]
LocationManager $name =
	(LocationManager)getSystemService(
		Context.LOCATION_SERVICE);
\end{lstlisting}
\caption{}
\label{fig:androidLocation}
\end{subfigure}
~
\begin{subfigure}[b]{0.3\textwidth}
	\begin{lstlisting}[basicstyle=\footnotesize\ttfamily]
Location.distanceBetween(
		$(Location).getLatitude(),
		$(Location).getLongitude(),
		$...);
\end{lstlisting}
\caption{}
\label{fig:androidLocation2}
\end{subfigure}
~
\begin{subfigure}[b]{0.3\textwidth}
	\begin{lstlisting}[basicstyle=\footnotesize\ttfamily]
try {
	$BODY$
} finally {
  $(RevWalk).release();
}
\end{lstlisting}
\caption{}
\label{fig:jgitWalk}
\end{subfigure}
~
\begin{subfigure}[b]{0.3\textwidth}
\begin{lstlisting}[basicstyle=\footnotesize\ttfamily]
try {
  Node $name=$methodInvoc();
  $BODY$
} finally {
  $(Transaction).finish();
}
\end{lstlisting}
\caption{}
\label{fig:neo4jTransaction2}
\end{subfigure}
~
\begin{subfigure}[b]{0.3\textwidth}
\begin{lstlisting}[basicstyle=\footnotesize\ttfamily]
ConnectionFactory factory =
	new ConnectionFactory();
$methodInvoc();
Connection connection =
	factory.newConnection();
\end{lstlisting}
\caption{}
\label{fig:nettyConnection}
\end{subfigure}
~
\begin{subfigure}[b]{0.3\textwidth}
\begin{lstlisting}[basicstyle=\footnotesize\ttfamily]
while ($(ModelNode) != null) {
	if ($(ModelNode) == limit)
		break;
	$ifstatement
	$(ModelNode)=$(ModelNode)
		.getParentModelNode();
}
\end{lstlisting}
\caption{}
\label{fig:rhinoNodeVisiting}
\end{subfigure}
~
\begin{subfigure}[b]{0.3\textwidth}
\begin{lstlisting}[basicstyle=\footnotesize\ttfamily]
Document doc=Jsoup.connect(URL).
	userAgent("Mozilla").
	header("Accept","text/html").
	get();
\end{lstlisting}
\caption{}
\label{fig:jsoupDocumentRetrieval}
\end{subfigure}
~
\begin{subfigure}[b]{0.3\textwidth}
\begin{lstlisting}[basicstyle=\footnotesize\ttfamily]
if ($(Connection) != null) {
	try {
		$(Connection).close();
	} catch (Exception ignore) { }
}
\end{lstlisting}
\caption{}
\label{fig:nettyConnection}
\end{subfigure}
~
\begin{subfigure}[b]{0.3\textwidth}
\begin{lstlisting}[basicstyle=\footnotesize\ttfamily]
Traverser traverser
	=$(Node).traverse();
for (Node $name : traverser) {
	$BODY$
}
\end{lstlisting}
\caption{}
\label{fig:neo4jTrasversal}
\end{subfigure}
~
\begin{subfigure}[b]{0.3\textwidth}
\begin{lstlisting}[basicstyle=\footnotesize\ttfamily]
Toast.makeText(this,
	$stringLit,Toast.LENGTH_SHORT)
		.show()
\end{lstlisting}
\caption{}
\label{fig:androidToast}
\end{subfigure}
~
\begin{subfigure}[b]{0.3\textwidth}
\begin{lstlisting}[basicstyle=\footnotesize\ttfamily]
try {
	Session session
		=HibernateUtil
			.currentSession();
	$BODY$
} catch (HibernateException e) {
	throw new DaoException(e);
}
\end{lstlisting}
\caption{}
\label{fig:hibernateSession}
\end{subfigure}
~
\begin{subfigure}[b]{0.3\textwidth}
\begin{lstlisting}[basicstyle=\footnotesize\ttfamily]
catch (HibernateException e) {
  if ($(Transaction) != null) {
    $(Transaction).rollback();
  }
  e.printStackTrace();
}
\end{lstlisting}
\caption{}
\label{fig:hibernateTransaction}
\end{subfigure}
~
\begin{subfigure}[b]{0.3\textwidth}
\begin{lstlisting}[basicstyle=\footnotesize\ttfamily]
FileSystem $name
	=FileSystem.get(
		$(Path).toUri(),conf);
\end{lstlisting}
\caption{}
\label{fig:hadoopFs}
\end{subfigure}
~
\begin{subfigure}[b]{0.3\textwidth}
\begin{lstlisting}[basicstyle=\footnotesize\ttfamily]
(token=$(XContentParser).nextToken()) 
	!= XContentParser.Token.END_OBJECT
\end{lstlisting}
\caption{}
\label{fig:luceneTokenizer}
\end{subfigure}

\caption{Top cross-project idioms for \textsc{Library} projects (\autoref{tbl:inprojects}). Here 
we include idioms that appear in the test set files. We rank them by
the number of distinct files they appear in and restrict into presenting
idioms that contain at least one library-specific (\ie API-specific) 
identifier. The special 
notation \lstinline+$(TypeName)+ denotes the presence of a variable
whose name is undefined. \lstinline+$BODY$+ denotes a user-defined
code block of one or more statements, \lstinline+$name+ a freely
defined (variable) name, \lstinline+$methodInvoc+ a single method
invocation statement and \lstinline+$ifstatement+ a single \lstinline+if+
statement. All the idioms have been automatically identifies by \haggis}
\label{fig:snippets}
\end{figure*}

\subsection{Top Idioms}
\autoref{fig:snippets} shows the top idioms mined in the \libraryDset
dataset, ranked by the number of files in the test sets where each idiom has 
appeared in. The reader will observe their immediate usefulness.
Some idioms capture how to retrieve or instantiate an object.
For example, in \autoref{fig:snippets}, the idiom \ref{fig:rabbitmqConnection} captures
the instantiation of a message channel in RabbitMQ,
\ref{fig:hadoopFs} retrieves a handle for the Hadoop 
file system, \ref{fig:elasticBuilder} builds a
\texttt{SearchSourceBuilder} in Elasticsearch and \ref{fig:jsoupDocumentRetrieval}
retrieves a URL using JSoup. Other idioms capture important transactional properties
of code: idiom \ref{fig:jgitWalk} uses properly the memory-hungry
\texttt{RevWalk} object in JGit and \ref{fig:neo4jTransaction2} is
a transaction idiom in Neo4J.
Other idioms capture common error handling, such as \ref{fig:neo4jTransactionFailure} for Neo4J
and \ref{fig:hibernateSession} for a Hibernate transaction.
Finally, some idioms capture common operations, such as closing a connection
in Netty (\ref{fig:nettyConnection}), traversing through the database nodes (\ref{fig:neo4jTrasversal}),
visiting all AST nodes in a JavaScript file in Rhino (\ref{fig:rhinoNodeVisiting})
and computing the distance between two locations (\ref{fig:androidLocation2})
in Android. The reader may observe that these idioms provide a 
meaningful set of coding patterns for each library, capturing semantically
consistent actions that a developer is likely to need when 
using these libraries.

In \autoref{fig:sampleIdioms} we present a small set of Java-related idioms
mined across all datasets. These idioms represent
frequently used code patterns that would be included by default in tools such as
Eclipse's SnipMatch \cite{eclipseSnipMatch} and IntelliJ's live
templates \cite{jetbrainsLiveTemplates}. Defining constants (\autoref{fig:stringConst}),
creating loggers (\autoref{fig:logger}) and iterating through an iterable (\autoref{fig:iterator}) are some of the most common
language-specific idioms in Java. All of these idioms have been 
automatically identified by \haggis.

\begin{figure}[t]
\footnotesize
\begin{center}
\begin{tabular}{cp{1.4cm}rrrrrr} \toprule
&\multirow{2}{*}{Name} & \multicolumn{2}{c}{Precision} & \multicolumn{2}{c}{Coverage} & \multicolumn{2}{c}{Avg Size} \\ 
& &\multicolumn{2}{c}{(\%)} & \multicolumn{2}{c}{(\%)} & \multicolumn{2}{c}{(\#Nodes)} \\ \midrule
&\haggis & 8.5&$\pm3.2$ & 23.5&$\pm 13.2$ & 15.0&$\pm 2.1 $ \\
& \multicolumn{3}{l}{\scriptsize{$n_{min}=5$, $c_{min}=2$}}\\
\parbox[t]{1mm}{\multirow{3}{*}{\rotatebox[origin=c]{90}{\libraryDset}}}&\haggis & 16.9&$\pm 10.1$  & 2.8 &$\pm 3.0 $ & 27.9&$\pm 8.63$ \\
& \multicolumn{3}{l}{\scriptsize{$n_{min}=20$, $c_{min}=25$}} \\
&\deckard & 0.9&$\pm 1.3$& 4.1&$\pm 5.24$ & 24.6&$\pm 15.0$\\  
& \multicolumn{3}{l}{\scriptsize{minToks=10, stride=2, sim=1}}\\\midrule
\parbox[t]{1mm}{\multirow{2}{*}{\rotatebox[origin=c]{90}{\projectsDset }}}&\haggis & 14.4&$\pm 9.4$  & 30.29&$\pm 12.5$ & 15.46&$\pm 3.1$ \\
& \multicolumn{3}{l}{\scriptsize{$n_{min}=5$, $c_{min}=2$}} \\
&\haggis & 29.9&$\pm 19.4$  & 3.1 &$\pm 2.6$ & 25.3&$\pm 3.5$ \\ 
&\multicolumn{3}{l}{\scriptsize{$n_{min}=20$, $c_{min}=25$}} \\\bottomrule
\end{tabular}
\end{center}
\normalsize 
\vspace{-1.5em}\caption{Average and standard deviation of performance in \libraryDset 
test set. Standard deviation across projects.}\vspace{-1em}
\label{tbl:deckard}
\end{figure}
We now quantitatively evaluate the mined idiom sets. \autoref{tbl:deckard}
shows idiom coverage, idiom set precision and the average size of the
matched idioms in the test sets of each dataset. We observe that \haggis
achieves better precision and coverage in \projectsDset. This is expected
since code idioms recur more often in a similar project rather than 
across disparate projects. This effect may be partially attributed to the 
small number of people working in a project and partially to project-specific
idioms. \autoref{tbl:deckard} also gives an indication of the trade-offs
we can achieve for different $c_{min}$ and $n_{min}$.

\subsection{Code Cloning vs Code Idioms}
Previously, we discussed that code idioms differ significantly from
code clones. We now show this by using a cutting-edge code clone detection
tool: \deckard \cite{jiang2007deckard} is a state-of-the-art tree-based
clone-detection tool that uses an intermediate vector representation for
detecting similarities. To extract code idioms from the code clone clusters that 
\deckard computes, we retrieve the maximal common subtree of each cluster,
ignoring patterns that are less that 50\% of the original size of the
tree.

We run \deckard with multiple parameters ($\text{stride}\in
\{0,2\}$, $\text{similarity}\in\{0.95,1.0\}$, $\text{minToks}\in\{10,20\}$) on the
validation set and picked the parameters that achieve the best combination
of precision and coverage.
\autoref{tbl:deckard} shows	 precision, coverage and average idiom
size (in number of nodes) of the patterns found through \deckard and
\haggis. \haggis found larger and higher coverage
idioms, since clones seldom recur across projects.
The differences in precision and coverage are statistically significant (paired $t$-test; $p < 0.001$).
We also note that the overlap in the patterns extracted by \deckard and \haggis  is small (less than 0.5\%).

It is important to note these results are not a criticism of 
\deckard---which \emph{is} a high-quality, state-of-the-art code clone detection 
tool---but rather, these results show that \emph{the task of code clone detection is
different from code idiom mining}: Code clone detection
is concerned with finding pieces of code that are not necessarily
frequent but are maximally identical. In contrast, idiom mining is not 
concerned with finding maximally identical pieces of code, 
but mining common tree fragments that trade-off between size and frequency.

\subsection{Extrinsic Evaluation of Mined Idioms}
Now, we evaluate \haggis extrinsically by computing coverage and
precision in the test sets of each dataset and the StackOverflow
question dataset \cite{bacchelli2013mining}, an extrinsic set of
highly idiomatic code snippets.
StackOverflow is a popular Q\&A site containing programming-related
questions and answers. When developers deem that their question or answer needs to
be clarified with code, they include a code snippet. These snippets are
representative of general development practice and are usually short,
concise and idiomatic, containing only essential pieces of code.
We first extract all code fragments in questions and answers tagged
as \textsf{java} or \textsf{android}, filtering only those that can
be parsed by Eclipse JDT \cite{eclipseJdt}. We further remove
snippets that contain less than 5 tokens. After this process, we have 
108,407 partial Java snippets. Then, we create a single set of
idioms, merging all those found in \libraryDset and removing any
idioms that have been seen in less than five files at the test portions
of \libraryDset. We end up with small but high precision set of idioms
across all APIs in \libraryDset.

\begin{figure}[t]
\begin{center}
\begin{tabular}{p{2cm}rr} \toprule
 Test Corpus & Coverage & Precision \\ \midrule 
 StackOverflow & 31\% & 67\% \\
\projectsDset & 22\% & 50\% \\\bottomrule
\end{tabular}
\end{center}
\normalsize
\vspace{-1.5em}\caption{Extrinsic evaluation of mined idioms. All idioms
were mined from \libraryDset.}\vspace{-1em}
\label{tbl:soEval}
\end{figure}

\begin{figure*}[t]
	\centering
	\includegraphics[width=0.9\textwidth]{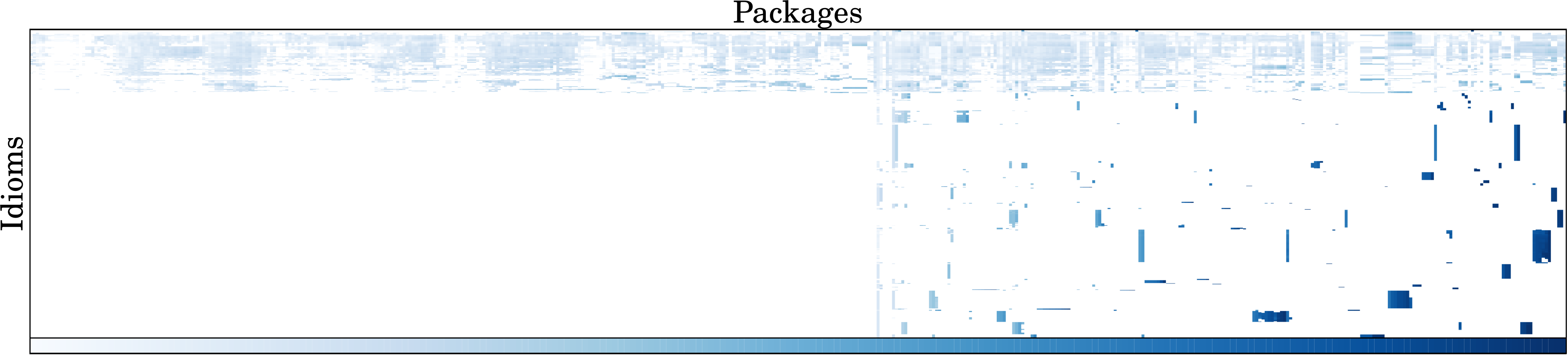}
	\caption{Lift between package imports and code idioms. A darker color
	signifies higher lift, \ie more common co-occurrence.
	Each row shows the ``spectrum'' of an idiom. Darker blue color shows
	higher correlation between a package and an idiom. 
	One can find idioms generic-language idioms (top) and others
	that are package-specific (dark blocks on the right).
	Idioms and packages are only shown for the \texttt{android.location},
	\texttt{android.net.wifi} and \texttt{org.hibernate} APIs for 
	brevity.
	}
	\label{fig:importPattern}\vspace{-1em}		
\end{figure*}

\autoref{tbl:soEval} shows precision and coverage of \haggis's
idioms comparing StackOverflow, \libraryDset and \projectsDset. Using
the \libraryDset idioms, we achieve a coverage of 31\% and a precision 
of 67\% on StackOverflow, compared to a much smaller precision and coverage
in \projectsDset. This shows that the mined idioms are more frequent
in StackOverflow than in a ``random'' set of projects. Since we expect that
StackOverflow snippets are more highly idiomatic than average projects'
source code, this provides strong indication that \haggis has mined
a set of meaningful idioms. 
We note that precision depends highly on the popularity of
\libraryDset's libraries. For example, because Android is one of the most
popular topics in StackOverflow, when we limit the mined idioms to those
found in the two Android libraries, \haggis achieves a precision
of 96.6\% at a coverage of 21\% in StackOverflow. 
This evaluation provides a strong indication that \haggis idioms
are widely used in development practice.

\subsection{Idioms and Code Libraries}
\label{sec:libraries}
Previously, we found code idioms across projects and libraries. As a final
evaluation of the mined code idioms' semantic consistency, we now show
that code idioms are highly correlated with the packages that are
imported by a Java file. We  merge
the idioms across our \libraryDset projects and visualize
the \emph{lift} among code idioms and \lstinline+import+ statements. Lift,
commonly used in association rule mining, measures how dependent the
co-appearance of two elements is. For each imported package $p$, 
we compute the lift score $l$ of the code idiom $t$ as
$l(p,t) = P(p,t)/ \left(P(p)P(t)\right)$
where $P(p)$ is the probability of importing package $p$, $P(t)$ is the probability
of the appearance of code idiom $t$ and $P(p,t)$ is the probability that package $p$ and 
idiom $t$ appear together. It can be seen that $l(p,t)$ is higher as
package $p$ and idiom $t$ are more correlated, \ie, their appearance is not
independent.

\autoref{fig:importPattern} shows a covariance-like matrix of the lift of
the top idioms and packages.
Here, we visualize the top 300 most frequent train set packages and their
highest correlating code idioms, along with the top 100 most frequent
idioms in \libraryDset. Each row represents a single code idiom and each column a
single package. On the top of \autoref{fig:importPattern} one can see
idioms that do not depend strongly on the package imports. These are
language-generic idioms (such as the exception handling idiom in \autoref{fig:stringConst})
and do not correlate significantly with any package.
We can also observe dark blocks of packages and idioms. Those 
represent library or project-specific idioms that co-appear frequently.
This provides additional evidence that \haggis finds meaningful 
idioms since, as expected, some idioms are common throughout
Java, while others are API or project-specific.

\boldpara{Suggesting idioms}
To further demonstrate the semantic consistency of the \haggis idioms,
we present a preliminary approach to suggesting idioms based on package imports.
We caution that our goal here is to develop an initial proof of concept,
not the best possible suggestion method.
First, we score each 
idiom $\mathcal{T}_i$ by computing $s(\mathcal{T}_i|\mathbb{I})=\max_{p\in \mathbb{I}}l(p,\mathcal{T}_i)$
where $\mathbb{I}$ is the set of all imported packages. We then 
return a ranked list $\mathbb{T}_{\mathbb{I}}=\left\{ \mathcal{T}_1, \mathcal{T}_2,\dots \right\}$
such that for all $i<j$, $s(\mathcal{T}_i,\mathbb{I}) > s(\mathcal{T}_j,\mathbb{I})$.
Additionally, we use a threshold $s_{th}$ to control the precision
of the returned suggestions, showing only those idioms $t_i$ that have
$s(\mathcal{T}_i,\mathbb{I}) > s_{th}$. Thus, we are only suggesting idioms where
the level of confidence is higher than $s_{th}$. It follows, that this
parameter controls suggestion frequency \ie the percent of the times where 
we present at least one code idiom.

\begin{figure}[t]
	\centering
		\includegraphics[width=0.9\columnwidth]{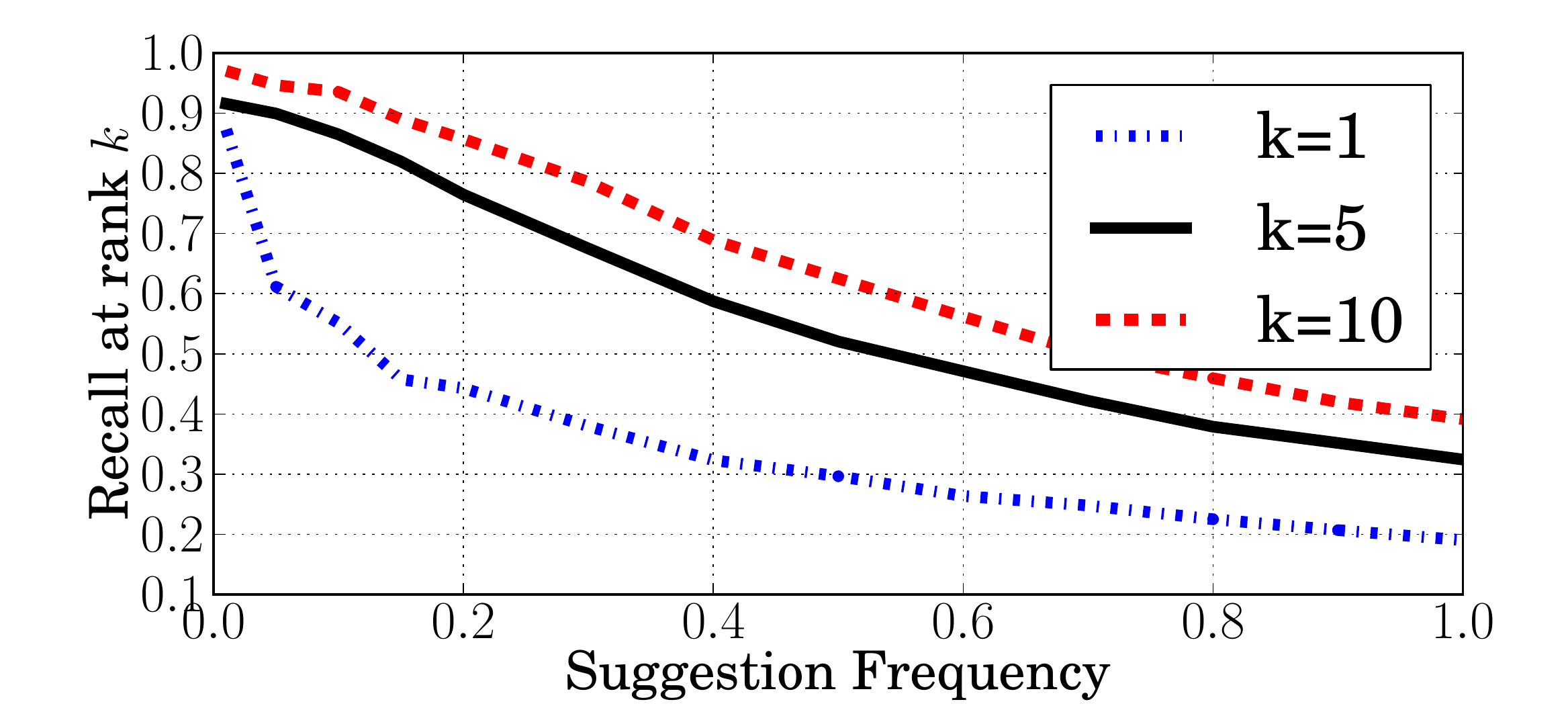}
	\caption{The recall at rank $k$ for code idiom suggestion.}
	\label{fig:idiomSuggestionRecall}	
	\vspace{-1.5em}
\end{figure}

To evaluate \haggis's idiom suggestions, we use the \libraryDset 
idioms mined from the train set and compute the recall-at-rank-$k$ on 
the \libraryDset's test set. Recall-at-rank-$k$ evaluates \haggis's
ability to return at least one code idiom for each test file.
\autoref{fig:idiomSuggestionRecall} shows that for suggestion frequency 
of 20\% we can achieve a recall of 76\% at rank $k=5$, meaning that in 
the top 5 results we return at least one relevant code idiom 76\% of the
time. This results shows the quality of the mined idioms, suggesting
that \haggis can provide a set of meaningful suggestions to a developer
by solely using the code's imports. Further improvements in suggestion performance
can be achieved by using more advanced classification methods, which
we leave to future work, which could eventually enable
an IDE side-pane that presents a list of suggested code idioms.

\section{Related Work}
Source code has been shown to be highly repetitive and non-unique 
\cite{gabel2010study} rendering NLP methods attractive for the analysis of source code.
\Ngram language models have been used
\cite{allamanis2013mining,hindle2012naturalness,nguyen2013statistical}
to improve code autocompletion performance, learn coding conventions
\cite{allamanis2014learning} and find syntax errors \cite{charlessyntax}.
Models of the tree structure of the code have also been studied 
with the aim of generating programs by example \cite{menon2013machine}
and modeling source code \cite{maddison2014structured}. However, none of this work has tried to extract 
non-sequential patterns in code or mine tree fragments. The only work
that we are aware of that uses language models for detecting textual patterns 
in code is Jacob and Tairas \cite{jacob2010code} that use \ngrams to 
autocomplete code templates.

Code clones \cite{cottrell2008semi,kamiya2002ccfinder,kapser2008cloning,kim2005empirical,li2006cp,roy2007survey,roy2009comparison}
are related to code idiom mining, since they aim to find highly
similar code, but not necessarily identical pieces of code. Code clone
detection using ASTs has also been studied extensively \cite{baxter1998clone,falke2008empirical,jiang2007deckard,koschke2006clone}. For a survey of clone detection methods, see Roy \etal \cite{roy2007survey,roy2009comparison}.
In contrast, as we noted in \autoref{sec:evaluation},
code idiom mining searches for frequent, rather than maximally identical
subtrees. It is worth noting that code clones have been found to 
have a positive effect on maintenance \cite{kapser2008cloning,kim2005empirical}.
Another related area is API mining 
\cite{acharya2007mining,holmes2006approximate,zhong2009mapo,wang2013mining}.
However, this area is also significantly different from code idiom mining because it
tries to mine sequences or graphs \cite{nguyen2009graph} of API method
calls, usually ignoring most features of the language. This difference should
be evident from the sample code idioms in \autoref{fig:snippets}.

Within the data mining literature, there has been 
a series of work on \emph{frequent tree mining} algorithms \cite{jimenez10frequent,termier02treefinder,zaki02efficiently,zaki05efficiently},
which focuses on finding subtrees that occur often in a database of trees.
However, as described in \autoref{sec:definition}, these have the difficulty
that frequent trees are not always interesting trees, a difficulty
which our probabilistic approach addresses in a principled way.
Finally, as described previously, Bayesian nonparametric
methods are a widely researched area in statistics and 
machine learning \cite{hjort2010bayesian,gershman2012tutorial,TehJor2010a,orbanz10}, which
have also found many applications in NLP 
\cite{Teh2006b,cohn2010inducing,post2009bayesian}.

\section{Discussion \& Conclusions}
In this paper, we presented \haggis, a system for automatically mining
high-quality code idioms. We found that code idioms appear in multiple settings:
some are project-specific, some are API-specific and some are 
language-specific. An interesting direction for future work is to study the reasons that 
code idioms arise in programming languages, APIs or projects and the
effects idioms have in the software engineering process. It could be that 
there are ``good'' and ``bad'' idioms. ``Good'' idioms may
arise as an additional abstraction layer over a programming language that
helps developers communicate more clearly the intention of their code.
``Bad'' idioms may compensate for deficiencies of a
programming language or an API. For example, one common Java
idiom mined by \haggis is a sequence of multiple catch statements.
This idiom is indeed due to Java's language design, that led Java language designers
to introduce a new 
``multi-catch'' statement in Java 7 \cite{javamulticatch}. However, other
idioms, such as the ubiquitous \lstinline{for(int i=0;i<n;i++)} 
cannot be considered a language limitation, but rather a useful and widely 
understandable code idiom. A more formal study of the difference
between these two types of idioms could be of significant interest.

\section*{Acknowledgments}
The authors would like to thank Jaroslav Fowkes, Sharon Goldwater and Mirella Lapata
for their insightful comments and suggestions. This work was supported
by Microsoft Research through its PhD Scholarship Programme and by the
Engineering and Physical Sciences Research Council [grant number EP/K024043/1].

%
\bibliographystyle{abbrvnat}
\bibliography{lit/patterns}

\end{document}